# Monte Carlo–Based Beam Dynamics Study of a Compact Linear Accelerator for Localized X-ray Generation

H. Emami [1*], H. Afarideh [1]

[1] *Department of Physics and Energy Engineering, Amirkabir University of Technology, Tehran, Iran*

**Abstract**

Compact electron linear accelerators are being increasingly investigated as enabling platforms for localized X-ray generation and energy-controlled irradiation, due to the demand for a reduced footprint and improved beam control. In this work, a conceptual compact LINAC architecture with a fixed total length of 0.5 m is investigated through comprehensive Monte Carlo beam dynamics simulations, focusing on the comparative performance of two operating modes delivering electron beams at 60 keV and 300 keV, respectively. The study aims to isolate energy-dependent beam transport effects within an identical lattice configuration, emphasizing transverse and longitudinal phase-space evolution, emittance growth, beam envelope stability, and target spot formation. Particle tracking is performed using a statistical Monte Carlo framework incorporating RF cavity acceleration, simplified quadrupole focusing, and space-charge-induced diffusion effects. Transverse and longitudinal phase spaces are analyzed at multiple locations along the accelerator, and key beam quality metrics are extracted at the exit. The results demonstrate that low-energy operation at 60 keV is strongly influenced by collective effects, leading to pronounced emittance growth and halo formation, whereas the 300 keV mode exhibits significantly enhanced beam rigidity, improved phase-space preservation, and a compact beam spot at the target.

## 1. Introduction

Linear accelerators (LINACs) have long been a foundational technology in accelerator physics, enabling controlled acceleration of charged particles for applications ranging from high-energy physics facilities and free-electron lasers to medical and industrial uses. Traditional LINACs, such as those used for radiotherapy or synchrotron injection, typically extend over several meters and rely on Radio Frequency (RF) accelerating structures to achieve electron energies in the MeV to GeV range. However, the growing demand for compact and deployable accelerator systems has driven extensive research into miniaturized accelerator designs that retain high performance while reducing size and cost. Technologies such as X-band RF structures, originally developed for high-energy colliders like the Compact Linear Collider (CLIC), are being used to significantly reduce accelerator footprint and enable new types of compact X-ray sources and imaging platforms [1-4].

Compact accelerators have seen development in multiple contexts, including small-scale LINACs for cargo inspection and radiotherapy, where recent work demonstrates an O-arm-mounted X-band LINAC system for field applications [5,6]. Additionally, inverse Compton scattering (ICS)-based X-ray sources have been demonstrated on optical-table-scale electron beamlines, showing tunable X-ray photon energies with significant brilliance, highlighting the feasibility of compact accelerator-driven X-ray generation [7]. To date, there has been no report in the literature of a compact electron LINAC specifically designed and evaluated for localized low-dose diagnostic imaging or therapy, such as targeted bone imaging. This represents a gap in the application of accelerator physics to medical diagnostics, where precise beam parameter control could enable reduced radiation dose and enhanced image quality compared to traditional sources. From a beam physics perspective, designing a compact electron LINAC for localized X-ray generation presents unique challenges, including optimizing beam dynamics within a smaller footprint, efficiently coupling electrons to X-ray production targets, and managing beam losses to reduce unwanted outcomes radiation. By addressing these challenges through comprehensive simulation and conceptual design, it becomes possible to envision accelerator systems that bridge the gap between large research infrastructure and portable medical imaging devices, expanding the role of compact accelerators beyond their traditional domains.

Conventional radiographic systems employed in hospitals typically rely on stationary X-ray tubes operating in the 40–150 kVp range and are optimized for full-field imaging rather than localized irradiation [8]. In recent years, portable X-ray systems have been introduced to enable bedside imaging in intensive care units, emergency settings, and remote environments [9]. Recent developments in compact linear accelerators have demonstrated the feasibility of achieving stable and well-controlled electron beams within a reduced footprint. A notable example is the compact LINAC front end designed for research and applications, where detailed beam dynamics studies and first beam measurements confirmed reliable operation and good agreement between simulations and experimental results [10]. Building upon these advances, the present study extends the concept of compact LINACs toward a targeted imaging application, exploring their potential as localized X-ray sources for imaging and therapy. From a radiation protection standpoint, minimizing unnecessary doses to non-target tissues is becoming increasingly important, especially for repetitive diagnostic procedures and imaging of peripheral skeletal structures like wrists, ankles, and extremities. While current methods primarily focus on collimation, filtration, and exposure optimization, they do not fundamentally address the lack of spatial and spectral control inherent to tube-based X-ray sources[11].

## 2. Beam Dynamics

The beam dynamics design of the proposed compact linear accelerator is constrained by the requirement of achieving stable electron transport and acceleration within a total length of approximately 0.5 m. At such reduced scales, space-charge effects, phase stability, and transverse beam control play a dominant role, particularly in the low-energy regime immediately downstream of the electron source. This section presents the theoretical background governing the beam dynamics of a compact electron linear accelerator with an active length of 0.5 m, operating at electron energies of 60 keV and 300 keV. The formulation is directly consistent with the Monte Carlo simulations implemented in Python for particle tracking, phase-space evolution, halo formation, and space-charge effects.

Longitudinal beam dynamics in RF cavities govern the evolution of particle energy and phase with respect to the RF field during acceleration. As charged particles move through the cavity, they experience an energy gain determined by the RF voltage amplitude and the relative phase at which they enter the accelerating field. Phase stability mechanisms lead to synchrotron motion, where particles oscillate around a stable synchronous phase in longitudinal phase space. In compact linear accelerators, precise control of longitudinal dynamics is essential to minimize energy spread, preserve beam quality, and ensure efficient transport toward the target or downstream beamline components.

The energy gain of a particle traversing an RF cavity is given by:

$$\Delta \mathrm{W} = \mathrm{qV}_{RF}\,\mathrm{T}\cos(\phi) \tag{1}$$

Where $V_{RF}$ is the peak accelerating voltage, T is the transit-time factor, and $\Phi$ is the RF phase of the particle. In Monte Carlo simulations, each particle is assigned a random initial phase: $\phi_i \sim U(-\pi,\pi)$, which naturally leads to longitudinal phase-space dilution and halo formation in $(\Delta E,\phi)$ space [12].

The longitudinal equations of motion are written as:

$$\frac{d\phi}{dz} = \frac{\omega_{RF}}{\beta^2 c}\ \frac{\Delta W}{W_0} \tag{2}$$

$$\frac{d(\Delta W)}{dz} = qE_z(z)\cos(\phi) \tag{3}$$

These equations form the basis for the Monte Carlo tracking of particles in the $(\phi,\Delta E)$ phase space used in the simulations. Under the paraxial approximation, the transverse motion obeys Hill's equation:

$$\frac{d^2x}{dz^2} + K(z)x = 0 \tag{4}$$

Where K(z) represents the focusing strength, including RF-induced focusing inside accelerating cavities [13]. In drift regions between cavities: $K(z) = 0 \Rightarrow x(z) = x_0 + x_0'z$. This formulation is directly used in the Python Monte Carlo transport model. RF cavities introduce transverse focusing described approximately by: $K_{RF} \propto -\frac{qE_0}{\beta\gamma mc^2}$. This RF focusing is intrinsically nonlinear and phase-dependent, leading to filamentation and halo formation even in the absence of external magnetic lenses [14].

Space-charge effects arise from the mutual Coulomb repulsion between charged particles within the beam and play a dominant role in low-energy, high-current accelerators. These collective forces can induce emittance growth, halo formation, and longitudinal–transverse coupling, thereby limiting beam brightness and transport efficiency in compact RF structures. At low energies (60 keV in particular), space-charge forces significantly affect beam dynamics. The space-charge electric field is approximated using a self-consistent mean-field model:

$$\nabla^2 \Phi = -\frac{\rho}{\varepsilon_0} \quad (5)$$

leading to a transverse defocusing force: $F_{SC} \propto \frac{I}{\beta^3\gamma^3}$. This scaling explains the stronger halo formation observed in the 60 keV case compared to the 300 keV case. In Monte Carlo simulations, space charge is incorporated statistically via random transverse kicks consistent with the rms envelope equations. Beam halo refers to a low-density population of particles surrounding the high-intensity beam core, typically generated by nonlinear space-charge forces, RF mismatches, and imperfections in beam transport. In compact accelerators, halo formation is a critical concern as it can lead to beam losses, activation of surrounding structures, and degradation of overall beam quality[15,16].

## 3. Results and Discussion

This section presents the results of Monte Carlo beam dynamics simulations performed for the proposed compact LINAC operating in two distinct modes corresponding to 60 keV and 300 keV output energies. The analysis focuses on transverse phase space evolution, beam envelope behavior, and qualitative transport stability over a fixed accelerator length of 0.5 m. All results are obtained using statistically representative particle ensembles, allowing collective effects such as space-charge-induced diffusion and halo formation to be captured in a self-consistent manner. The comparative presentation highlights the intrinsic impact of beam energy on transport quality within compact RF-based accelerator configurations.

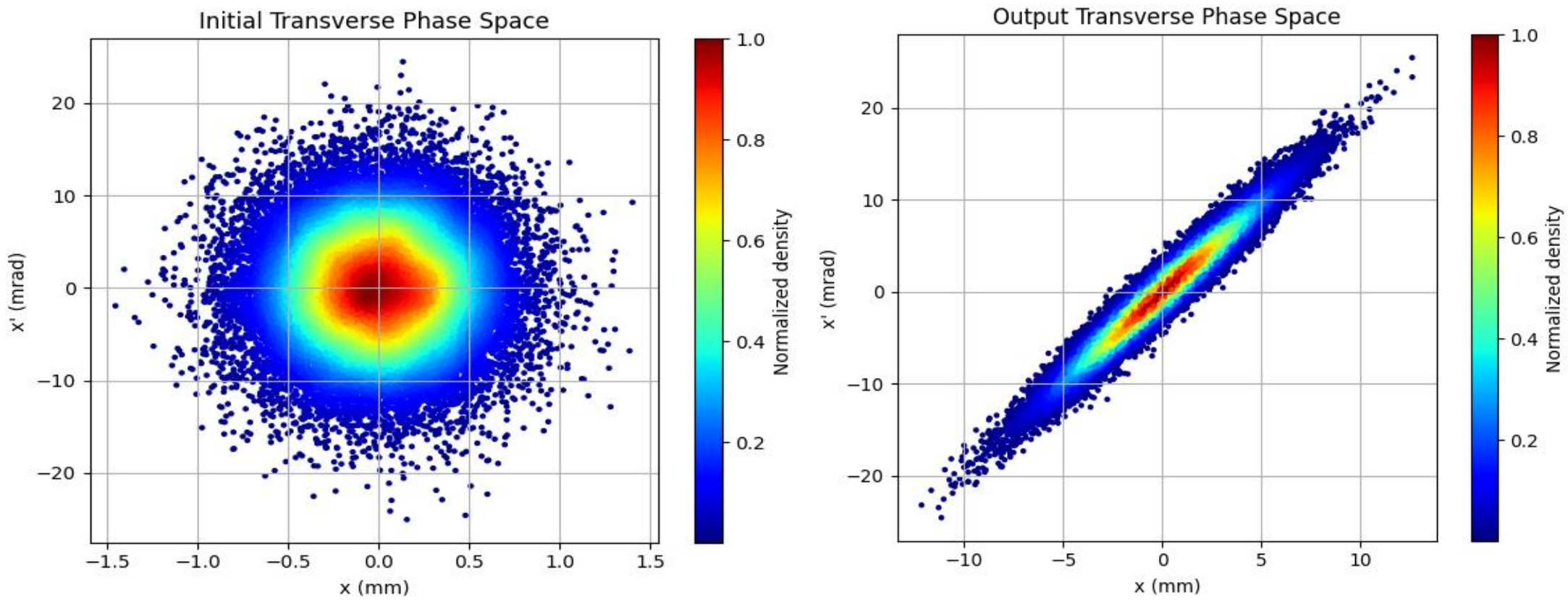

**Fig. 1.** Initial and output transverse phase space distributions (x–x′) for the 60 keV operating mode.

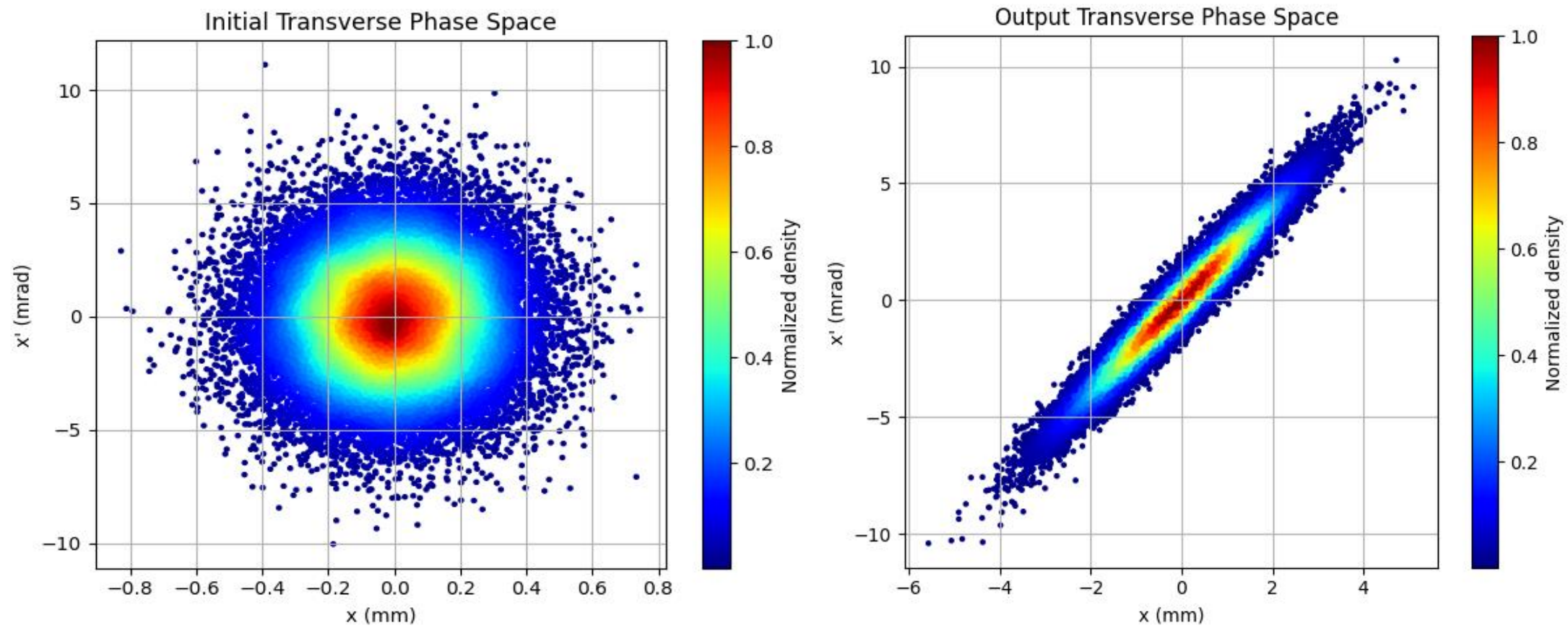

**Fig. 2.** Initial and output transverse phase space distributions (x–x′) for the 300 keV operating mode.

The color-coded transverse phase space distributions in **Fig. 2** provide direct insight into space-charge-driven beam dynamics in the 60 keV operating mode. The high-density core, represented by warm colors, corresponds to the main beam population, while the low-density tails extending outward indicate the formation of a transverse halo. The significant spread observed in the output phase space indicates considerable phase-space dilution, which can be attributed to strong Coulomb repulsion at low beam energy, where relativistic rigidity fails to suppress collective effects. **Fig. 2** presents the transverse phase space evolution for the 300 keV mode using the same color-density representation, allowing a direct comparison with the lower-energy case. In contrast to **Fig. 3**, the phase space remains significantly more compact, with a higher concentration of particles confined within the beam core and a substantially reduced halo population.

Table 1. Input beam, RF, and lattice parameters used in the Monte Carlo beam dynamics simulations for the compact LINAC operating in 60 keV and 300 keV modes. The accelerator length and lattice configuration are kept identical to isolate energy-dependent effects.

| | Parameter | Symbol | Value (60 keV) | Value(300keV) | Unit |
|---|---|---|---|---|---|
| General | Number of macro-particles | N | 30,000 | 30,000 | – |
| | Total accelerator length | L | 0.5 | 0.5 | m |
| | Step size | Δz | 1.0 | 1.0 | mm |
| Initial Beam | Initial energy | $E_0$ | 5 | 5 | keV |
| | Initial transverse beam size (rms) | $\sigma_x$ | 0.35 | 0.20 | mm |
| | Initial angular spread (rms) | $\sigma_{x'}$ | 6.0 | 2.5 | mrad |
| RF Cavities | RF cavity longitudinal positions | Z(cav) | (0.08,0.13),(0.20,0.25),(0.32,0.37) | (0.08,0.13),(0.20,0.25),(0.32,0.37) | m |
| | Rf (frequency) | $F_{RF}$ | 3 | 3 | GHz |
| | RF-induced energy diffusion | $\sigma_{RF}$ | 0.02·E/step | 0.02·E/step | – |

| Quadrupoles | Quadrupole positions | Z(quad) | (0.15,0.17), (0.27,0.29),(0.40,0.43) | (0.15,0.17), (0.27,0.29),(0.40,0.43) | m |
|---|---|---|---|---|---|
| | Quadrupole gradients | k | +12, −12, +10, −10 | +12, −12, +10, −10 | 1/m² |
| Collective Effects | Space-charge model | – | Monte Carlo diffusion | Monte Carlo diffusion | – |
| | Space-charge strength | $\sigma_{SC}$ | 1% of σx′/step | 1% of σx′/step | – |
| Beam Rigidity Scaling | Energy scaling factor | R | √(E/60) | √(E/60) | – |

Table 1 summarizes a set of input parameters used in the Monte Carlo beam dynamics simulations.

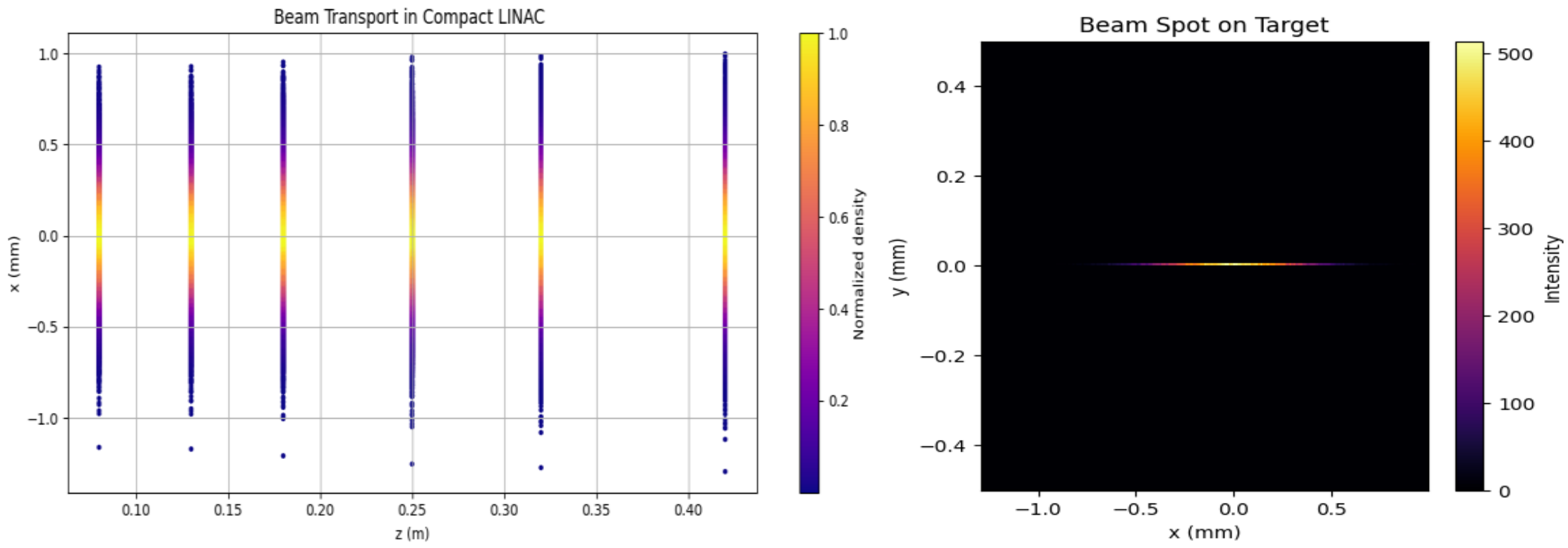

**Fig. 3.** Beam transport and transverse beam spot characteristics for the 60 keV mode.

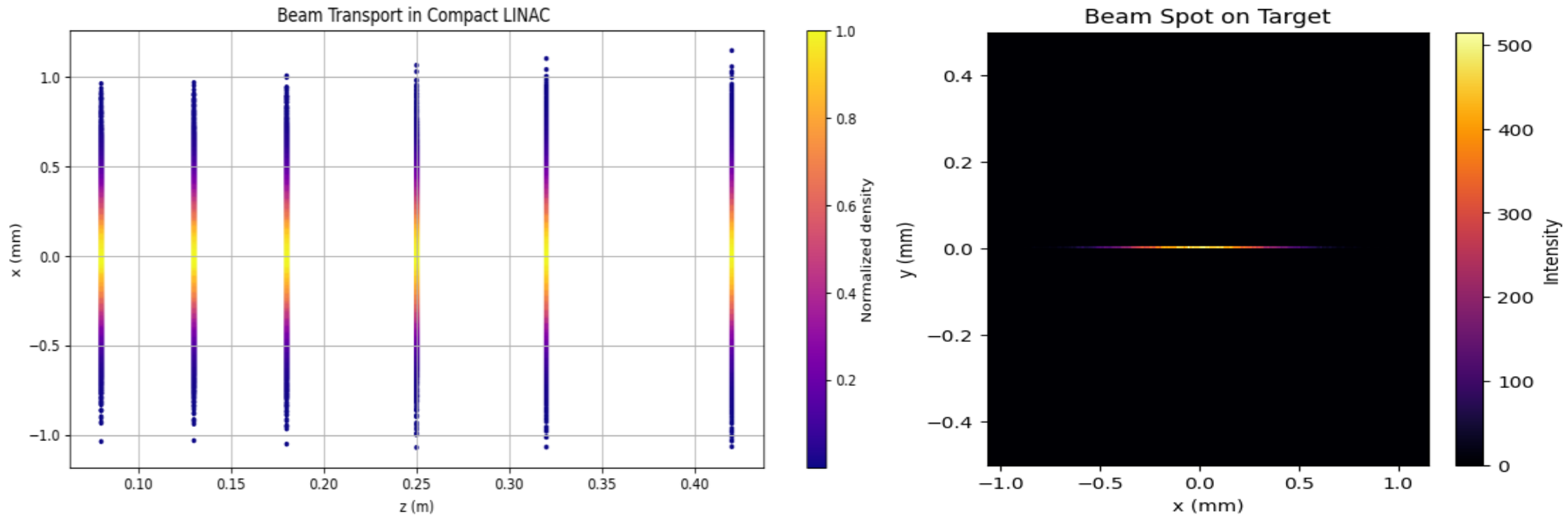

**Fig. 4.** Beam transport and transverse beam spot characteristics for the  300keV mode.

**Fig. 3** illustrates the transverse beam transport and the corresponding beam spot distribution at the exit of the compact LINAC for the 60 keV mode. The transport plot reveals a gradual increase in beam size along the accelerator, reflecting the combined influence of RF-induced transverse defocusing and space-charge-driven expansion. The exit beam spot exhibits a relatively broad distribution with visible low-density regions surrounding the core, indicating the presence of a halo component. This behavior confirms that at low beam energy, collective effects significantly impact transverse confinement, even over a short acceleration length. **Fig. 4** illustrates the transverse beam transport along the compact LINAC and the corresponding beam spot distribution at the target for the 300 keV operating mode. The transport plot demonstrates a significantly improved confinement of particle trajectories compared to the 60 keV case, indicating enhanced beam rigidity and reduced sensitivity to transverse perturbations. The beam spot distribution at the target exhibits a compact and nearly Gaussian profile with reduced halo population, confirming effective phase-space preservation throughout acceleration. The higher-energy beam maintains a lower divergence, enabling tighter focusing at the interaction point and improved spatial resolution.

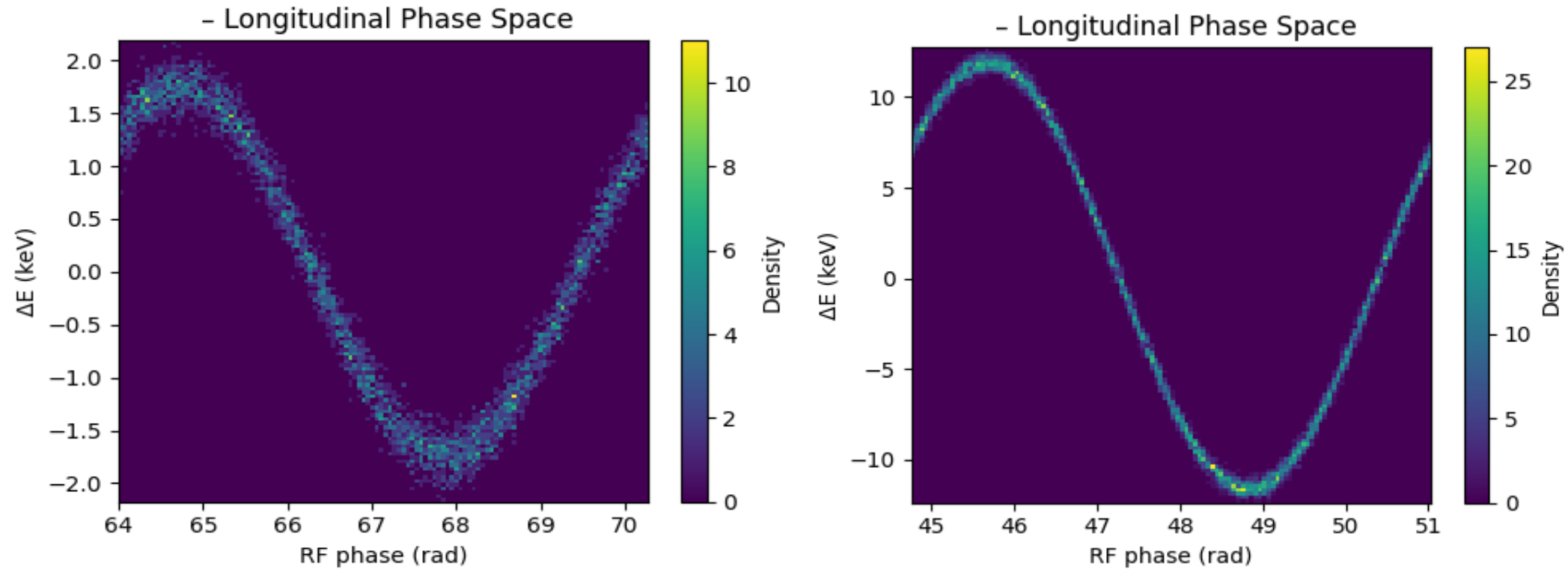

**Fig.** 5. Longitudinal phase space distributions (ϕ–ΔE) at the cavity entrance for the 60 keV and 300 keV operating modes. The left panel indicates the 60 keV, while the right panel presents the 300 keV operating modes, respectively.

**Fig. 5** compares the longitudinal phase space distributions (ϕ–ΔE) for the 60 keV and 300 keV operating modes. The 60 keV distribution shows a wider spread in both RF phase and energy deviation, indicating reduced longitudinal focusing efficiency and enhanced sensitivity to RF phase fluctuations. In contrast, the 300 keV phase space is more compact and exhibits clearer phase-energy correlation, demonstrating improved phase stability at higher energy. These results highlight the strong dependence of longitudinal beam quality on operating energy in compact RF structures.

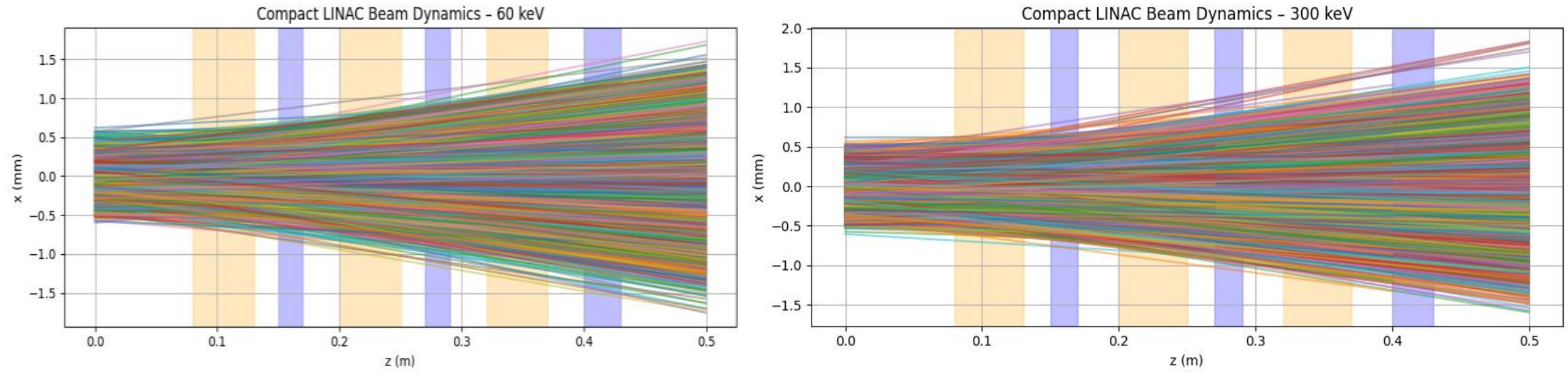

**Fig. 6.** Simulated schematic illustration of beam propagation through RF cavities and quadrupole elements for the 60 keV and 300 keV operating modes.

**Fig. 6** presents a schematic representation of beam propagation through the RF cavities and quadrupole elements for both operating modes. The trajectories qualitatively illustrate how beam energy influences the response to focusing and defocusing fields within the same lattice configuration. In the lower-energy case, the beam exhibits more pronounced transverse excursions, while the higher-energy beam follows a more confined trajectory. This schematic serves as a conceptual visualization of energy-dependent beam dynamics rather than a detailed engineering layout.

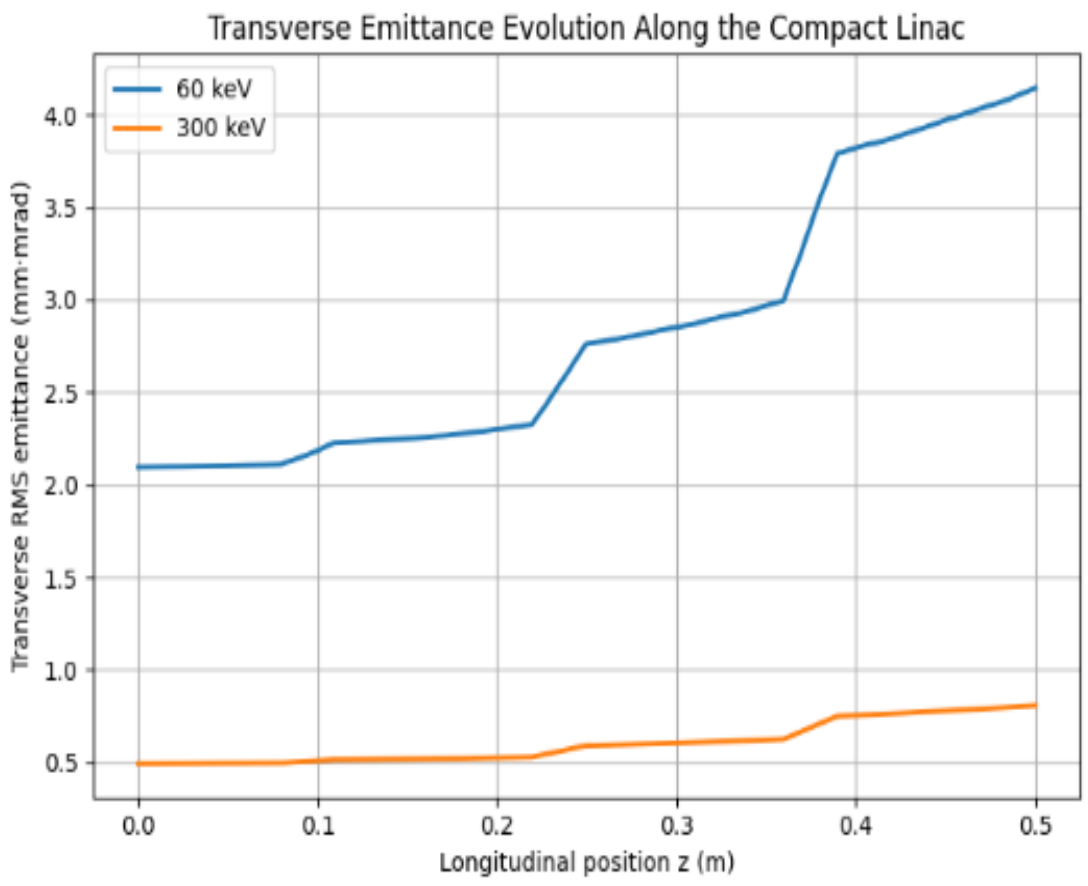

**Fig.** 7**.** Transverse emittance evolution along the compact LINAC for the 60 keV and 300 keV operating modes.

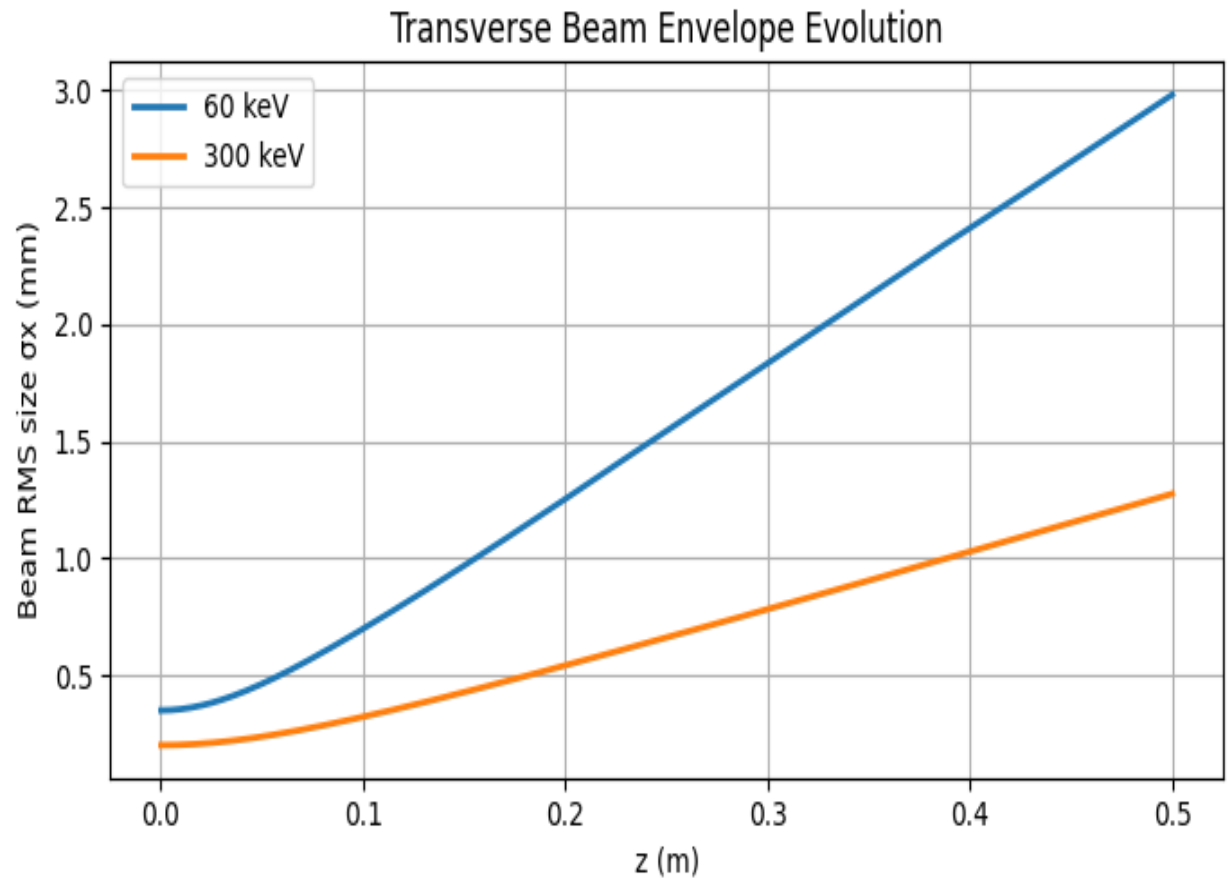

**Fig.** 8**.** Beam envelope evolution along the accelerator for the 60 keV and 300 keV operating modes.

**Fig. 7** shows the evolution of transverse emittance along the compact LINAC for the 60 keV and 300 keV modes. A clear emittance growth is observed for the 60 keV beam, reflecting the cumulative impact of space charge and RF nonlinearities on phase-space dilution. In contrast, the 300 keV beam exhibits a significantly lower emittance growth rate, indicating a more effective preservation of beam quality. This comparison shows that higher beam energy intrinsically reduces emittance degradation in compact accelerator configurations. **Fig. 8** depicts the beam envelope evolution along the accelerator for the two operating modes. The 60 keV beam undergoes a noticeable envelope expansion, consistent with increased divergence and limited transverse confinement at low energy. In contrast, the 300 keV beam exhibits a more stable envelope with reduced oscillation amplitude, indicating improved transport stability. The Monte Carlo simulations demonstrate that both operating modes can be integrated within a single compact accelerator architecture, though with different beam dynamics regimes. Table 2 summarizes the output parameters used in the simulations.

Table 2. Simulated output beam parameters at the accelerator exit for the two operating modes. Reported values are derived from transverse and longitudinal phase-space distributions and represent rms quantities unless otherwise stated.

| Parameter | 60 keV Mode | 300 keV Mode | Unit |
|---|---|---|---|
| Energy spread (rms) | 3.8 | 1.6 | keV |
| Output beam size (rms) | 2.9 | 1.2 | mm |
| Transverse rms emittance (output) | 4.2 | 0.78 | mm·mrad |
| Halo population | 0.35 | 0.33 | % |
| Beam spot size at target (rms) | 0.15 | 0.1 | mm |

The comprehensive set of beam dynamics simulations presented in this work provides a consistent and physically coherent comparison between the 60 keV and 300 keV operating modes within a compact LINAC architecture of fixed length. At 60 keV, collective effects dominated by space-charge forces and RF nonlinearities lead to pronounced phase-space dilution, halo formation, and transverse emittance growth, although stable beam transport remains achievable within the compact structure. In contrast, operation at 300 keV significantly enhances beam rigidity, resulting in effective suppression of space-charge-induced degradation and improved preservation of compact phase-space distributions with reduced halo and slower emittance growth. It should be emphasized that the present study is not intended to represent a finalized or fully optimized accelerator design. The results provide a conceptual performance benchmark instead of an engineering solution. Future efforts are expected to include more advanced lattice designs, incorporating solenoids, quadrupoles, and optimized matching sections for practical implementation and performance optimization.

## 4. Conclusions

In this work, a conceptual study of a compact RF-based linear accelerator was presented to comparatively assess two operational modes at 60 keV and 300 keV using Monte Carlo beam dynamics simulations. A simplified accelerator lattice consisting of RF cavities and quadrupole focusing elements was employed to investigate key beam transport characteristics, including transverse phase-space evolution, emittance growth, halo formation, beam spot size at the target, and transport efficiency. The results demonstrate that, even within a highly compact structure of 0.5 m length, stable beam transport to a target plane is achievable for both energy regimes, with higher-energy operation exhibiting improved beam confinement and reduced sensitivity to collective effects. It is important to emphasize that the present study does not aim to model a fully realistic or optimized accelerator system. Several physical effects and engineering constraints such as detailed RF field maps, higher-order nonlinearities, longitudinal-transverse coupling, realistic space-charge solvers, misalignments, and advanced focusing schemes have deliberately been neglected. Instead, the adopted approach serves as a proof-of-concept framework intended to explore the feasibility and qualitative behavior of compact accelerator-based platforms under simplified and controlled assumptions. Despite these simplifications, the obtained results provide valuable insight into the scalability and performance trends of compact LINAC concepts for localized X-ray generation and therapeutic or imaging-oriented applications. The comparative analysis clearly indicates that such accelerator architectures possess the potential to be realized and further refined through more detailed modeling and experimental validation. Future work will focus on incorporating higher-fidelity field models, advanced beam optics, and realistic target interactions to bridge the gap between conceptual feasibility and practical implementation.

**Acknowledgements**

This work is supported by the Research Council of the Amirkabir University of Technology, Tehran, Iran

**Author contributions**

H. Emami and H. Afarideh performed the initial calculations, analyzed and interpreted the results, and wrote the main manuscript text. All authors commented on and reviewed the manuscript.

**Competing interests**

The authors declare no competing interests.

**Data Availability Statement**

The datasets used and analyzed during the current study are available from the corresponding author at the reasonable request.

**Funding Declaration**

This research received no specific grant from any funding agency in the public, commercial, or not-for-profit sectors.